\documentstyle[aps,prl,psfig,floats]{revtex}

\newcommand{\be}{\begin{equation}}
\newcommand{\ee}{\end{equation}}
\def\(#1){(\ref{#1})}
\newcommand{\bea}{\begin{eqnarray}}
\newcommand{\eea}{\end{eqnarray}}

\newcommand{\deriv}[1]{{\partial\over\partial{#1}}}
\newcommand{\lav}{\left\langle}
\newcommand{\rav}{\right\rangle}
\newcommand{\eg}{{\it e.g.\,\,}}

\newcommand{\av}[1]{\left\langle #1 \right\rangle}

\newcommand{\tw}{t_{\rm w}}
\newcommand{\dt}{t\!-\!t_{\rm w}}
\newcommand{\twlim}{t_{\rm w}\to\infty}
\newcommand{\tlim}{t\to\infty}
\newcommand{\mw}{m_{\rm w}}
\newcommand{\Ew}{E_{\rm w}}
\newcommand{\distrm}{\sigma(m|E)}
\newcommand{\trapmean}{\overline{m}(E)}

\newcommand{\trapvar}{\Delta^2(E)}
\newcommand{\globalmbr}[1]{\left\langle m(#1) \right\rangle}
\newcommand{\D}{\displaystyle}
\newcommand{\Emprobc}{P(E,m,t|E_{\rm w},m_{\rm w},t_{\rm w})}
\newcommand{\Ct}{\tilde{C}}
\newcommand{\Rt}{\tilde{\chi}}
\newcommand{\C}{C(t-t_{\rm w})}
\newcommand{\R}{\chi(t-t_{\rm w})}
\newcommand{\CC}{C(t,t_{\rm w})}
\newcommand{\RR}{\chi(t,t_{\rm w})}
\newcommand{\XX}{X(t,t_{\rm w})}
\newcommand{\teff}{T_{\rm eff}}
\newcommand{\tg}{T_{\rm g}}

\begin{document}

\draft 

\twocolumn[\hsize\textwidth\columnwidth\hsize\csname @twocolumnfalse\endcsname

\title{Observable-dependence of fluctuation-dissipation relations and
effective temperatures}

\author{Suzanne Fielding$^1$~\cite{present_address} and Peter Sollich$^2$}

\address{
$^1$Department of Physics and Astronomy, University of Edinburgh, Mayfield Road, Edinburgh, EH9 3JZ, United Kingdom\\
$^2$Department of Mathematics, King's College London, Strand, London, WC2R 2LS, United Kingdom
}

\maketitle

\begin{abstract}

We study the non-equilibrium version of the fluctuation-dissipation
theorem (FDT) within the glass phase of Bouchaud's trap model. We
incorporate into the model an arbitrary observable $m$ and obtain its
correlation and response functions in closed form. A limiting
non-equilibrium FDT plot (of correlator vs response) is approached at
long times for most choices of $m$, with energy-temperature FDT a
notable exception. In contrast to standard mean field models, however,
the shape of the plot depends nontrivially on the observable, and its
slope varies continuously even though there is a single scaling of
relaxation times with age. Non-equilibrium FDT plots can therefore not
be used to define a meaningful effective temperature $T_{\rm eff}$ in
this model. Consequences for the wider applicability of an FDT-derived
$T_{\rm eff}$ are discussed.

\end{abstract}


\pacs{PACS: 05.20.-y; 05.40.-a; 05.70.Ln; 64.70.Pf}]

Glassy systems relax extremely slowly at low temperatures. They
therefore remain far from equilibrium on very long time scales, and
exhibit {\em ageing}~\cite{BouCugKurMez98}: the time scale of response
to an external perturbation increases with the age (time since
preparation) $\tw$ of the system. Time translational invariance (TTI)
and the equilibrium fluctuation-dissipation theorem~\cite{Reichl80}
(FDT) relating correlation and response functions break
down.

Let $\CC=\lav m(t)m(\tw)\rav -\lav m(t) \rav \lav m(\tw) \rav$ be the
autocorrelation function for an observable $m$, $R(t,\tw)={\delta
\lav m(t)\rav}/{\delta h(\tw)}|_{h=0}$ the linear response of $m(t)$
to a small impulse
in its conjugate field $h$ at time $\tw$, and $\RR=\int_{\tw}^t\!
dt'\, R(t,t')$ the corresponding response
to a field step $h(t)=h\Theta(\dt)$. In {\em
equilibrium}, $\CC=\C$ by TTI (similarly for $R$ and $\chi$), and the FDT
reads $-\deriv{\tw} \R= R(t,\tw) = \frac{1}{T}\deriv{\tw}\C$, with $T$ the
thermodynamic temperature (we set $k_{\rm B}=1$). A parametric ``FDT
plot'' of $\chi$ vs.\ $C$ is thus a straight line of slope $-1/T$. In
the ageing case, the
violation of FDT can be measured by a
fluctuation-dissipation ratio (FDR), $\XX$, defined
through~\cite{CugKur93,CugKur94}
\be
\label{eqn:non_eq_fdt}
-\deriv{\tw} \RR= R(t,\tw) = \frac{\XX}{T}\deriv{\tw}\CC
\ee
In equilibrium, due to TTI, the derivatives $\deriv{\tw}$ in the FDT
can be replaced by $-\deriv{t}$; in the non-equilibrium
definition~(\ref{eqn:non_eq_fdt}) this would make less sense since
only the $\tw$-derivative of $\RR$ is directly related to the impulse
response $R(t,\tw)$. Values of $X$ different from unity mark a
violation of FDT, and can persist even in the limit of long times,
indicating strongly non-equilibrium behavior even though
one-time quantities such as entropy and average energy may evolve
infinitesimally slowly.

Remarkably, the FDR for several {\em mean field}
models~\cite{CugKur93,CugKur94} assumes a special form at long times:
Taking $\tw\to\infty$ at constant $C=C(t,\tw)$, $X(t,\tw)\to X(C)$
becomes a (nontrivial) function of the single argument $C$. If the
equal-time correlator $C(t,t)$ also approaches a constant $C_0$ for
$t\to\infty$, it follows that $\chi(t,\tw)=\int_{C(t,\tw)}^{C_0}
\!dC\, X(C)/T$. Graphically, this limiting non-equilibrium FDT
relation is obtained by plotting $\chi$ vs $C$ for increasingly large
times; from the slope $-X(C)/T$ of the limit plot, an {\em effective
temperature}~\cite{CugKurPel97} can be defined as $\teff(C)=T/X(C)$.
In such a scenario, where $\chi(t,\tw)$ and $C(t,\tw)$ become simple
functions of each other, either $\tw$ or $t$ can be used as the curve
parameter for the FDT plot; the latter is the conventional
prescription~\cite{CugKur93,CugKur94}. In general, however, the
definition~(\ref{eqn:non_eq_fdt}) ensures a slope of $-\XX/T$ for a
parametric $\chi$-$C$ plot {\em only} if $\tw$ is used as the parameter,
with $t$ being fixed.

In the most general ageing scenario, a system displays dynamics on
several characteristic time scales, one of which may remain finite as
$\twlim$, while the others diverge with $\tw$.
If, due to their different functional dependence on $\tw$, 
these time scales become
infinitely separated 
as $\twlim$, they form a set of
distinct ``time sectors''; in mean field, $\teff(C)$ can then be shown
to be {\em constant} within each such sector~\cite{CugKur94}. In the short
time sector ($\dt=O(1)$), where $C(t,\tw)$ decays from $C_0$ to some
plateau value,
one
generically has quasi-equilibrium with $\teff=T$, giving an initial
straight line with slope $-1/T$ in the FDT plot. The further decay of
$C$ (on ageing time scales $\dt$ that grow with $\tw$) gives rise to
one of three characteristic shapes: (i) In models which statically
show one step replica symmetry breaking (RSB), \eg the spherical
$p$-spin model~\cite{CugKur93}, there is only one ageing time sector
and the FDT plot exhibits a second straight line, with $\teff>T$. (ii)
In models of coarsening and domain growth, \eg the $O(n)$ model at
large $n$, this second straight line is flat, and hence
$\teff=\infty$~\cite{CugDea95}. (iii) In models with an infinite
hierarchy of time sectors (and infinite step RSB in the statics, \eg
the SK model) the FDT plot is instead a continuous
curve~\cite{CugKur94}.

$\teff$ has been interpreted as a time scale dependent non-equilibrium
temperature, and within mean field has been shown to display many of
the properties associated with a thermodynamic
temperature~\cite{CugKurPel97}. For example (within a given
time sector), it is the reading which would be shown by a thermometer
tuned to respond on that time scale. Furthermore---and of crucial
importance to its interpretation as a temperature---it is independent
of the observable $m$ used to construct the FDT
plot~\cite{CugKurPel97}. 

While the above picture is well established in mean field, its status
in non-mean field models is less obvious. To check its validity,
one must demonstrate that a) a limiting FDT plot
exists and that b) it gives effective temperatures that are
independent of the observable-field pair used to calculate $C$ and
$\chi$. Depending on whether there are one or many ageing time sectors
one then expects FDT plots similar to those for mean field
systems. 

Encouragingly, molecular dynamics (MD) and Monte Carlo (MC)
simulations of binary Lennard-Jones mixtures~\cite{KobBar00}, as well
as MC simulations of frustrated lattice gases~\cite{AreRicSta00}
(which loosely model structural glasses, whose phenomenology is
similar to that of the $p$-spin model) show a limiting plot of type
(i). MC simulations of the Ising model with conserved and
non-conserved order parameter in dimension $d=2$ and 3 show a plot of
type (ii)~\cite{Barrat98}. And MC simulations of the Edwards-Anderson
model in $d=3$, 4~\cite{MarParRicRui98} give a plot of type (iii). The
majority of existing studies, however, do not show that $\teff$ is
independent of observable (notable exceptions
are~\cite{KobBar00,AreRicSta00}) since they consider just one
observable-field pair.

In this work, therefore, we analyse a model where FDT plots can be
calculated analytically for {\em arbitrary} observables, allowing us to
investigate in detail whether the simple mean field picture applies.
Trap models~\cite{Bouchaud92,MonBou96,RinMaaBou00} are obvious
candidates for such a study. Popular as alternatives to the
microscopic spin models discussed above, they capture ageing within a
simplified single particle description. The simplest such
model~\cite{Bouchaud92} comprises an ensemble of uncoupled
particles exploring a spatially unstructured landscape of (free)
energy traps by thermal activation. The tops of the traps are at a
common energy level and their depths $E$
have a `prior' distribution $\rho(E)$ ($E>0$).
A particle in a trap of depth $E$ escapes on 
a time scale $\tau(E)=\tau_0\exp({E}/{T})$ and hops into
another trap, the depth of which is drawn at random from
$\rho(E)$.  The probability, $P(E,t)$, of finding a
randomly chosen particle in a trap of depth $E$ at time $t$ thus
obeys
\be
\label{eqn:trap_eom}
(\partial/\partial t)P(E,t)=-\tau^{-1}(E)P(E,t)+Y(t)\rho(E)
\end{equation}
in which the first (second) term on the RHS represents hops out of
(into) traps of depth $E$, and $Y(t)$ $=$ $\av{\tau^{-1}(E)}_{P(E,t)}$ is
the average hopping rate. The solution of~(\ref{eqn:trap_eom}) for
initial condition $P_0(E)$ is
\bea 
\label{eqn:trap_solution}
P(E,t)&=&P_0(E)e^{-{t}/{\tau(E)}}\nonumber\\
      & &{}+{}\rho(E)\int_0^{t}\!\!
dt' \,Y(t') e^{-(t-t')/\tau(E)}
\eea
from which $Y(t)$ has to be determined self-consistently.  For the
specific choice of prior distribution
$\rho(E)\sim\exp\left(-{E}/{\tg}\right)$, the model shows a glass
transition at a temperature $\tg$.  This can be seen as follows.  At a
temperature $T$, the equilibrium Boltzmann state (if it exists) is
$P_{\rm eq}(E)\propto\tau(E)\rho(E)\propto\exp\left({E}/{T}\right)
\exp\left(-{E}/{\tg}\right)$. For temperatures $T\le \tg$ this is
unnormalizable, and cannot exist; the lifetime averaged over the
prior, $\av{\tau}_{\rho}$, is infinite.  Following a quench to
$T\le\tg$, the system never reaches a steady state, but instead ages: in
the limit $\tw\to\infty$, $P(E,\tw)$ is concentrated entirely in traps
of lifetime $\tau=O(\tw)$.  The model thus has just one characteristic
time scale, which grows linearly with the age $\tw$. (In contrast, for
$T>\tg$ all relaxation processes occur on time scales $O(1)$.) In what
follows, we rescale all energies such that $\tg=1$, and also set
$\tau_0=1$.

To study FDT we extend the model 
by assigning to each trap,
in
addition to its depth $E$, a value for an (arbitrary) observable $m$;
by analogy with spin models we refer to $m$ as magnetization.  The
trap population is then characterized by the joint prior distribution
$\distrm\rho(E)$, where $\distrm$ is the distribution of $m$ across
traps of given fixed energy $E$. We focus on the non-equilibrium
dynamics after a quench at $t=0$ from $T=\infty$ to $T<1$; the initial
condition
is thus
$P_0(E,m)=\distrm\rho(E)$.  The subsequent evolution is governed by
\bea
\label{eqn:eomwithmh}
\lefteqn{(\partial/\partial t)P(E,m,t)=}\nonumber\\
& &\ \ \ \ -\tau^{-1}(E,m)P(E,m,t)+
Y(t)\rho(E)\distrm
\eea
where the activation times are modified by 
a small field $h$
conjugate to $m$ 
as
$\tau(E,m)=\tau(E)\exp\left(mh/T\right)$. Other 
choices of
$\tau(E,m)$ that maintain detailed balance are
possible~\cite{RinMaaBou00,BouDea95}; we adopt this particular one
because, in the spirit of the unperturbed model, it ensures that the
jump rate between any two states depends only on the initial state,
and not the final one.

The autocorrelation function of $m$ (for $h=0$) is
\bea
\CC&=&\int \!\! dm\ d\mw\ dE\ d\Ew\ m \, \mw \,P(\Ew,\mw,\tw)
\nonumber\\
   & &\left[\Emprobc-P(E,m,t)\right]
\label{eqn:correarly}
\end{eqnarray}
in which $\Emprobc$ is the probability that a particle with
magnetization $\mw$ and energy $\Ew$ at time $\tw$ subsequently has
$m$ and $E$ at time $t$. This obeys
\bea
\lefteqn{\Emprobc =}\nonumber\\
        & &\delta(m-\mw)\delta(E-\Ew) e^{-(t-t')/\tau(\Ew)}P(\Ew,\mw,\tw)\nonumber\\
        & &{}+{}\int_{\tw}^{t}\!\!dt' \, \tau^{-1}(\Ew)
e^{-(t'-\tw)/\tau(\Ew)}P(E,t-t')\distrm
\label{eqn:propagator}
\eea
The terms corresponds to the particle not having hopped at all since
$\tw$ (first term) or having first hopped at $t'$ (second term) into
another trap; after a hop the particle evolves as if ``reset'' to time
zero since it selects its new trap from the prior distribution
$\sigma(m|E)\rho(E)$, which describes the initial state of the system.
We also have $P(E,m,t)=\distrm P(E,t)$ since at zero field the
unperturbed dynamics is recovered. Substituting these relations
into~(\ref{eqn:correarly}) and integrating over $m$ and $\mw$, we
obtain an exact integral expression for $C$. This is expressible as
the sum of two components: The first depends only on the mean
$\trapmean$ of the fixed energy distribution $\distrm$ and the second
only on its variance, $\trapvar$.

To find the corresponding response function, we proved 
\be
T\deriv{\tw}\RR=\deriv{t}\CC +
\deriv{t}\globalmbr{t}\globalmbr{\tw}
\label{eqn:H3}
\end{equation}
in which $\lav m(t)\rav=\lav\trapmean\rav_{P(E,t)}$ is the global mean
of $m$. Equation~(\ref{eqn:H3}) generalizes the results
of~\cite{BouDea95,SasNem99} to the case of non-zero means $\lav
m\rav$; it is exact for any Markov process in which the effect of the
field on the transition rate between any two states depends only on
the initial state and not the final one. Substituting our expression
for $C$ into~(\ref{eqn:H3}), and integrating over $\tw$ with the
boundary condition $\chi(\tw,\tw)=0$, we find a closed expression for
$\chi$. This again comprises two components which depend separately on
the mean and variance of $\distrm$. For convenience, we rescale
the field $h\to Th$, thus absorbing a factor $1/T$ into the definition
of the response function. In this way, the slope of the FDT plot of
$\chi$ vs $C$ becomes $-X=-T/\teff$ ($=-1$ in equilibrium).

Using our exact expressions, we numerically calculated $C$ and $\chi$
for a number of different distributions $\distrm$, each specified by
given functional forms of $\trapmean$ and $\trapvar$. Each form
effectively corresponds to a distinct physical identity of the
observable $m$. For example $\distrm=\delta(m-E)$ implies $m=E$,
in which case $h=-\delta T/T$. For simplicity, we confined ourselves
to distributions either of zero mean (but non-zero variance) or of
zero variance (but non-zero mean). As expected, we find that 
the decay of $C$ (and growth of $\chi$) depends
not only on the time interval $\dt$, but also explicitly on the
waiting time $\tw$, consistent with the non-equilibrium breakdown of
TTI. For our generic observables $m$, the equal-time correlator
$C(t,t)$ can also depend strongly on $t$; this is in contrast to the
spin models often considered in mean field studies, where
$C(t,t)$ is automatically normalized. While from~(\ref{eqn:non_eq_fdt}) a
plot of $\chi$ vs $C$, with $t$ fixed and $\tw$ as the
parameter, still has slope $-X$ (with $t$ as parameter this
would no longer be guaranteed {\em a priori}), the variation of $C(t,t)$
will prevent a limit plot from being approached as we increase $t$.
We therefore normalize and plot 
$\Rt=\chi/C(t,t)$ vs $\Ct=C/C(t,t)$; the normalization factor $C(t,t)$ is
independent of $\tw$ and the same for both axes, ensuring that the
slope remains $-X$ as desired.

For the first class of observables (with zero mean $\trapmean$),
we considered a
variance $\trapvar=\exp(En/T)$ for various values of the exponent $n$,
generalizing the results of~\cite{BouDea95} for $n=0$. (For the weaker
power law dependence $\trapvar=E^{2p}$ we can show analytically that
the limiting FDT plot reduces to that for $n=0$.)  The equal-time
correlator is then
%
$C(t,t)=\int\! dE\ P(E,t)\exp\left({nE}/{T}\right)$.
%
For large $t$, its scaling behavior can be found using
%
\[
\renewcommand{\arraystretch}{1.2}
P(E,t) \sim \left\{
\begin{array}{lll} 
\D t^{T-1}\exp(E/T-E)  &\mbox{for} & \tau(E)=e^{E/T} \ll t \\
\D t^{T} \exp(-E) &\mbox{for} &  \tau(E)=e^{E/T} \gg t. \\
\end{array}
\right.
\]
\begin{figure}
\centerline{\psfig{figure=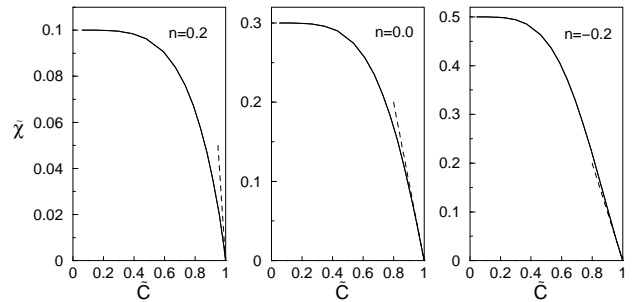,width=8.15cm}}
\caption{
  FDT plots of $\tilde{\chi}$ vs $\tilde{C}$ for a distribution
  $\distrm$ of variance $\exp(nE/T)$ (but zero mean) for
  $n=0.2,\,0.0,\,-0.2$; $T=0.3$. For each $n$ data are shown
  for times $t=10^6,\,10^7$; these
  are indistinguishable, confirming that the limiting FDT plot has
  been attained. Dashed line: The predicted asymptote
$\tilde{\chi}=1-\tilde{C}$ for
  $\tlim$ and $\tilde{C}\to 1$.
\label{fig:var}
}
\end{figure}
(which follows from~(\ref{eqn:trap_solution}), using $Y(t)\sim
t^{T-1}$~\cite{Bouchaud92,MonBou96}). For $n<T-1$, the integral for
$C(t,t)$ converges in a range $E=O(1)$, giving $C(t,t)\sim t^{T-1}$:
The equal-time correlator is sensitive only to shallow traps (the
population of which depletes in time as $t^{T-1}$ due to ageing), and
we thus expect the two-time correlator to decay on time scales
$\dt=O(1)$, probing only quasi-equilibrium behavior. In contrast, for
$T-1<n<T$ the integrand for $C(t,t)$ has most of its weight at
energies corresponding to $\tau(E)=O(t)$, yielding $C(t,t)\sim t^{n}$.
For such $n$, the two-time correlator will decay on ageing time scales
$\dt=O(\tw)$, and we expect strong violation of equilibrium FDT.  (The
regime $n>T$ is meaningless: It gives $C(t,t)=\infty\, \forall\, t$.) 

Our numerical results
confirm the above
predictions. For values of $n<T-1$ the normalized FDT plot approaches
a straight line of equilibrium slope $-1$ at long times $\tlim$. In
the regime $T-1<n<T$ (fig.~\ref{fig:var}), equilibrium FDT is
strongly violated, as expected; a limiting {\em non-equilibrium} FDT
plot is nevertheless approached at long times. This can be proved
analytically by showing that $\tilde C$ and $\tilde\chi$ share the
same scaling variable $(\dt)/\tw$ in this limit. The slope of each
plot varies continuously with $\tilde C$.
In contrast to mean field, this is not due to an infinite hierarchy of
time sectors; the variation is in fact continuous across the single
time sector $\dt=O(\tw)$. More seriously, different observables give
nontrivially different plots: at values of $\tilde C$ corresponding to
a fixed $(\dt)/\tw$, the slopes $-X$ are {\em not} independent of $n$.

\begin{figure}
\centerline{\psfig{figure=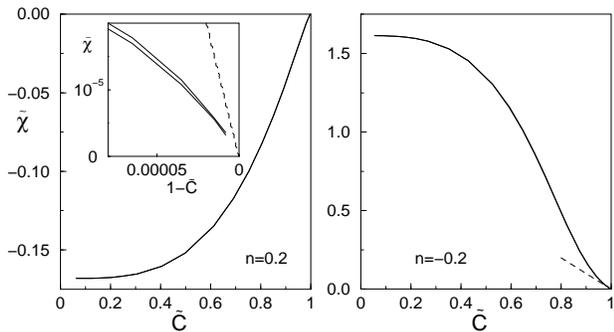,width=8.15cm}}
\caption{FDT plots of $\Rt$ vs $\Ct$ for a distribution with
  mean $\exp(nE/2T)$ (but zero variance), for $n=0.2,\,-0.2$; $T=0.3$.
  Curves are shown for times $t=10^6,\,10^7$,
  but are indistinguishable except for the zoom-inset on the left hand
(upper curve: $t=10^7$). Dashed line: The predicted asymptote
$\tilde{\chi}=1-\tilde{C}$ for $\tlim$, $\tilde{C}\to 1$. 
\label{fig:mean}
}
\end{figure}

Next we considered a fixed energy distribution $\distrm$ with zero
variance and mean $\trapmean=\exp(En/2T)$. Here too, we found that for
$n<T-1$ the limiting FDT plot is of equilibrium form, while for
$T-1<n<T$ (fig.~\ref{fig:mean}) a non-equilibrium FDT plot is
approached as $\tlim$, with a shape dependent (now very obviously) on
the observable $m$. Finally, we considered a power law dependence of
the mean, $\trapmean=E^p$; $p=1$ corresponds to energy-temperature
FDT. Interestingly, in this case no limiting plot exists. This is
because the amplitude of the correlator remains finite as $\tlim$,
while the response function, for any fixed value of $\tilde{C}$,
diverges as $\ln t$.

In summary, we have studied FDT in the glass phase of the trap model,
for a broad class of observables $m$. For observables which indeed
probe the ageing regime, equilibrium FDT is violated; in most cases, a
limiting non-equilibrium FDT plot is approached at long times. In
contrast to mean field, this plot depends strongly on the observable.
It furthermore has a slope which varies continuously across the single
time sector $\dt=O(\tw)$. This shows that in this simple, paradigmatic
model of glassy dynamics, the mean field concept of an FDT-derived
effective temperature $\teff$ cannot be applied.  One may dismiss the
trap model as too abstract for this to be of general relevance; but by
placing traps onto a $d$-dimensional lattice and allowing hops only to
neighbouring traps, a more `physical' model (of diffusion in the
presence of disorder) can be obtained. Recent work~\cite{RinMaaBou00}
shows the scaling of correlation functions to be unaffected by this
modification; since~(\ref{eqn:H3}) would also continue to hold, our
main results should be qualitatively unchanged.

Could the idea of a $\teff$ derived from FDT nevertheless be rescued,
in this and other non-mean field models? Consider first the
non-uniqueness of our limiting FDT plots. One could argue that in
order to probe an inherent $\teff$ characterizing the non-equilibrium
dynamics, the statistical properties of the observable must not change
significantly across the phase space regions visited during ageing.
(In coarsening models, similar arguments have been used to exclude
observables correlated with the order parameter~\cite{Barrat98}.)
Since in the trap model the typical trap depth $E$ increases without
bound for $t\to\infty$, a `neutral' observable would presumably
require that $\trapmean$, $\trapvar\to$ const.\ as $E\to\infty$; with
this restriction we indeed find a unique limiting FDT plot. Its slope
$-X$ will still vary continuously with $\dt$, however, excluding the
link to a thermodynamically meaningful $\teff$. Similarly `rounded'
FDT plots have recently been found in coarsening models at
criticality~\cite{GodLuc00b}; the limiting value $-X_\infty$ of the
slope for $C\to 0$ was there shown to be a universal amplitude ratio.
It is possible that at least this $X_\infty$ could define a sensible
$\teff$, and in fact all our limiting FDT plots above share a common
value $X_\infty=0$.  In conclusion, there is at least one class of
`critical-like' models (which includes the trap
model~\cite{trap_critical}) for which a non-equilibrium $\teff$ can be
defined at best from the asymptotic FDT $X_\infty$, and possibly only
for sufficiently neutral observables. It remains an open challenge to
characterize this class more fully, and to delineate it from those
models which {\em appear} to exhibit mean field-like behavior in
spite of their non-mean field
nature\cite{KobBar00,AreRicSta00,Barrat98}.

{\bf Acknowledgements}: We thank J.\ P.\ Bouchaud, M.\ E.\ Cates, M.\ 
R.\ Evans and F.\ Ritort for helpful suggestions, and EPSRC for
financial support (SMF).


\begin{thebibliography}{10}

\vspace*{-0.5cm}

\bibitem[*]{present_address}
Present address: Dept.\ of Physics and Astronomy $\&$ Polymer IRC,
  University of Leeds, Leeds, LS2 9JT, U.K.

\bibitem{BouCugKurMez98}
J.~P. Bouchaud, L.~F. Cugliandolo, J. Kurchan, and M. M{\'{e}}zard,  in {\em
  Spin glasses and random fields}, edited by A.~P. Young (World Scientific,
  Singapore, 1998).

\bibitem{Reichl80}
L.~E. Reichl, {\em A modern course in statistical physics} (University of Texas
  Press, Austin, 1980).

\bibitem{CugKur93}
L.~F. Cugliandolo and J. Kurchan, Phys.\ Rev.\ Lett. {\bf 71},  173  (1993).

\bibitem{CugKur94}
L.~F. Cugliandolo and J. Kurchan, J.\ Phys.\ A {\bf 27},  5749  (1994).

\bibitem{CugKurPel97}
L.~F. Cugliandolo, J. Kurchan, and L. Peliti, Phys.\ Rev.\ E {\bf 55},  3898
  (1997).

\bibitem{CugDea95}
L.~F. Cugliandolo and D.~S. Dean, J.\ Phys.\ A {\bf 28},  4213  (1995).

\bibitem{KobBar00}
W. Kob and J.~L. Barrat, Eur.\ Phys.\ J.\ B {\bf 13},  319  (2000).

\bibitem{AreRicSta00}
J.~J. Arenzon, F. Ricci-Tersenghi, and D.~A. Stariolo, Phys.\ Rev.\ E {\bf 62},
   5978  (2000).

\bibitem{Barrat98}
A. Barrat, Phys.\ Rev.\ E {\bf 57},  3629  (1998).

\bibitem{MarParRicRui98}
E. Marinari, G. Parisi, F. Ricci-Tersenghi, and J.~J. Ruiz-Lorenzo, J.\ Phys.\
  A {\bf 31},  2611  (1998).

\bibitem{Bouchaud92}
J.~P. Bouchaud, J.\ Phys.\ (France)\ I {\bf 2},  1705  (1992).

\bibitem{MonBou96}
C. Monthus and J.~P. Bouchaud, J.\ Phys.\ A {\bf 29},  3847  (1996).

\bibitem{RinMaaBou00}
B. Rinn, P. Maass, and J.-P. Bouchaud, Phys.\ Rev.\ Lett. {\bf 84},  5403
  (2000).

\bibitem{BouDea95}
J.~P. Bouchaud and D.~S. Dean, J.\ Phys.\ (France)\ I {\bf 5},  265  (1995).

\bibitem{SasNem99}
M. Sasaki and K. Nemoto, J.\ Phys.\ Soc.\ Jpn. {\bf 68},  1148  (1999).

\bibitem{GodLuc00b}
C. Godreche and J.~M. Luck, J.\ Phys.\ A {\bf 33},  9141  (2000),
  cond-mat/0001264.

\bibitem{trap_critical}
The trap model may be `effectively critical' since it
  describes the dynamics of the Random Energy Model around the freezing
  threshold. See G.~J.~M. Koper and H.~J. Hilhorst, Europhys.\ Lett. {\bf 3},  1213  (1987).

\end{thebibliography}

\end{document}